\documentclass[conference]{IEEEtran}

\usepackage{amsmath}
\usepackage{amssymb}
\usepackage{amsfonts}
\usepackage{algorithmic}
\usepackage{graphicx}
\usepackage{textcomp}
\usepackage{xcolor}
\usepackage[style=ieee]{biblatex}
\usepackage{arydshln}
\usepackage{cprotect}
\bibliography{references}

\begin{document}

\title{Inline Detection of Domain Generation Algorithms with Context-Sensitive Word Embeddings}

\author{
    \IEEEauthorblockN{Joewie J. Koh\IEEEauthorrefmark{1}\IEEEauthorrefmark{2} and Barton Rhodes\IEEEauthorrefmark{1}}
    \IEEEauthorblockA{\IEEEauthorrefmark{1}Optfit LLC, Denver, Colorado\\
    \{joewie, barton\}@optfit.ai}
    \IEEEauthorblockA{\IEEEauthorrefmark{2}Georgia Institute of Technology, Atlanta, Georgia\\
    joewie@ml.engineering}
}

\maketitle

\begin{abstract}
Domain generation algorithms (DGAs) are frequently employed by malware to generate domains used for connecting to command-and-control (C2) servers. Recent work in DGA detection leveraged deep learning architectures like convolutional neural networks (CNNs) and character-level long short-term memory networks (LSTMs) to classify domains. However, these classifiers perform poorly with wordlist-based DGA families, which generate domains by pseudorandomly concatenating dictionary words. We propose a novel approach that combines context-sensitive word embeddings with a simple fully-connected classifier to perform classification of domains based on word-level information. The word embeddings were pre-trained on a large unrelated corpus and left frozen during the training on domain data. The resulting small number of trainable parameters enabled extremely short training durations, while the transfer of language knowledge stored in the representations allowed for high-performing models with small training datasets. We show that this architecture reliably outperformed existing techniques on wordlist-based DGA families with just 30 DGA training examples and achieved state-of-the-art performance with around 100 DGA training examples, all while requiring an order of magnitude less time to train compared to current techniques. Of special note is the technique's performance on the \textit{matsnu} DGA: the classifier attained a 89.5\% detection rate with a 1:1,000 false positive rate (FPR) after training on only 30 examples of the DGA domains, and a 91.2\% detection rate with a 1:10,000 FPR after 90 examples. Considering that some of these DGAs have wordlists of several hundred words, our results demonstrate that this technique does not rely on the classifier learning the DGA wordlists. Instead, the classifier is able to learn the semantic signatures of the wordlist-based DGA families.
\end{abstract}


\section{Introduction}

Various families of malware require communication with command-and-control (C2) servers in order to receive instructions, exfiltrate collected intelligence, or engage in other malicious endeavors. Since hardcoded rendezvous addresses are susceptible to trivial countermeasures once discovered, malware often employs domain generation algorithms (DGAs) to generate a large number of pseudorandom domains against which connection attempts are made. This behavior is difficult to defeat due to the asymmetric nature of the threat: security professionals must deny access to the entire set of domains that can be generated by the DGA in order to contain the attack, while the malware author needs to control just a single domain to maintain communication \cite{plohmann2016}.

The problem of identifying malicious domains has been widely pursued in the past decade, primarily with techniques relying on manual feature engineering \cite{zhauniarovich2018}. The first featureless approach to the problem using hidden Markov models (HMMs) was presented by Antonakakis et al. (2012) \cite{antonakakis2012}. In the first application of deep learning to the problem of classifying DGA domains, Woodbridge, Anderson, Ahuja, and Grant (2016) showed that character-level long short-term memory networks (LSTMs) were extremely effective at identifying DGA domains with the exception of wordlist-based DGAs \cite{woodbridge2016}. Other authors have built upon this work in \cite{yu2017, lison2017, mac2017, tran2018, yu2018, choudhary2018, spaulding2018} with both convolutional neural networks (CNNs) and LSTMs, but improvements in detecting wordlist-based DGAs were limited.

The problem of identifying wordlist-based DGA domains is challenging as these domains do not necessarily look random. For example, a domain such as \mbox{\textit{dhlpcscshdrvpcpp.com}} (generated by the \verb|ramnit| DGA) is considerably more suspicious than a domain like \mbox{\textit{middleapple.net}} (generated by the \verb|suppobox| DGA). The reason these techniques had poor performance on wordlist-based DGAs was likely that they were primarily trained to detect random-looking strings.

However, some recent progress has been made towards solving the issue of poor detection performance on wordlist-based DGAs domains. Yang, Liu, Zhai, and Dai (2018) proposed a random forest classifier that used manually-extracted features such as word frequency, part-of-speech tags, and word correlations for classifying wordlist-based DGAs \cite{yang2018}. Pereira, Coleman, Yu, De Cock, and Nascimento (2018) designed a graph-based approach for learning the wordlists used by wordlist-based DGAs \cite{pereira2018}. While successful, their approach required a large number of training examples relative to the size of the DGA's wordlist because they identified candidate words by looking for common substrings between domains. Curtin, Gardner, Grzonkowski, Kleymenov, and Mosquera (2018) proposed the \textit{smashword score} as a measure of domain resemblance to English words, and developed a model that utilized WHOIS information to attain good performance on high-scoring families \cite{curtin2018}.

This paper presents a novel real-time technique for inline detection of wordlist-based DGA domains. The main idea behind our technique was that the validity of the context inherent in domains could contain sufficient information with which to identify DGA domains: unlike legitimate domains, which usually contain multiple words that are contextually valid together, DGA domains consist of words that do not typically appear in the same context. We thus utilized context-sensitive word embeddings to infer the aforementioned context validity. Combining said embeddings with a simple fully-connected classifier, we were able to perform classification of domains based on word-level information. To reduce training durations, we used pre-trained word embeddings and froze the embedding layers during the training on domain data to minimize the number of trainable parameters. The use of word embeddings that were pre-trained on a large unrelated corpus also enabled significant transfer learning to take place, which allowed for models that required minimal training data to achieve good performance. This approach reliably outperformed existing methods on wordlist-based DGA families with just 30 DGA training examples and achieved state-of-the-art performance with around 100 DGA training examples. Moreover, training durations were on the scale of minutes, compared to some current techniques in the literature that required multiple hours of training.

The technique does not rely on manual feature engineering or the classifier learning the DGA wordlists, nor does it require augmentation with external information, unlike \cite{yang2018, pereira2018, curtin2018}. Instead, the classifier learns to identify the semantic signatures of the wordlist-based DGA families, demonstrated by the fact that the classifier was able to identify DGA domains even when their constituent words had never been observed in training. We believe that this is the first work that applied word-level embeddings or transfer learning to DGA detection. While not in the realm of DGA detection, Fang, Peng, Liu, and Huang (2018) used word vectors of Structured Query Language (SQL) tokens and an LSTM model to detect SQL injection attacks \cite{fang2018}. To the best of our knowledge, this was the only previous application of natural language word-level embeddings in the detection of malicious network activity.

\section{Background}

In natural language processing (NLP), pre-trained word embeddings are widely used to alleviate the need for large amounts of task-specific training data. The approach takes advantage of unsupervised training on a large unrelated corpus and transfers learned knowledge to downstream models in the form of pre-trained word embeddings. A critical component of our technique is the context-sensitive word embedding, which was a relatively recent development in NLP. It refers to a word embedding that is able to differentiate between disparate word contexts and provide distinct representations for polysemes\footnote{Polysemes are words that have several meanings.}. For instance, the word ``bow'' in ``bow and arrow'' is unlike the word ``bow'' in ``bow towards the audience''. Two popular context-sensitive word embeddings are context vectors (CoVe) \cite{mccann2017} and Embeddings from Language Models (ELMo) \cite{peters2018}. We opted to use ELMo in our work as it was shown to outperform CoVe in most settings \cite{peters2018}.

The authors of ELMo refer to it as a \textit{deep contextualized} representation: \textit{deep} being that the embedding is dependent on all the internal layers of the bidirectional LSTM instead of just the final layer, and \textit{contextualized} being that the representation of each word is dependent on the entire input context. An additional essential feature of the ELMo representation is that the representations are character-based and thus able to adapt to out-of-vocabulary tokens by relying on morphological cues. Remarkable performance improvements were demonstrated for NLP tasks when previous state-of-the-art embeddings were replaced with ELMo. The authors speculated that contextual representations encode information that benefit NLP tasks, and that the inclusion of this contextual information in ELMo is the driving force behind the performance improvements in the tasks. We will show that the contextual information in such embeddings is useful even outside of traditional NLP tasks.

\section{Method}

We performed two preprocessing steps on the input data, with the first consisting of removing top-level domains (TLDs) when present. Removal of TLDs was performed as the extra information was unnecessary to achieve good performance with our model. Moreover, it is relatively trivial for malware authors to change the TLDs used by their DGAs, and a model that is able to identify DGA domains without requiring the TLD information would be less susceptible to such a countermeasure.

The second step of preprocessing involved splitting strings into their word-level components. Since ELMo is a word embedding, it performs best if its inputs are tokenized. While there are multiple techniques for doing this, the Python package \textit{wordninja}\footnote{https://github.com/keredson/wordninja} was simple to integrate with our codebase and performed sufficiently well for our purposes. The \textit{wordninja} implementation assumes that unigram frequencies follow a Zipfian distribution for computing the probability-maximizing split using dynamic programming. As the algorithm is linear in both time and memory, its utilization did not have performance implications for our technique. Alternatives include the \textit{WordSegment}\footnote{https://github.com/grantjenks/python-wordsegment} Python package and the \textit{WordSegmentationTM}\footnote{https://github.com/wolfgarbe/WordSegmentationTM} .NET library.

\begin{figure}[htbp]
\centering{\includegraphics[width=0.9\linewidth]{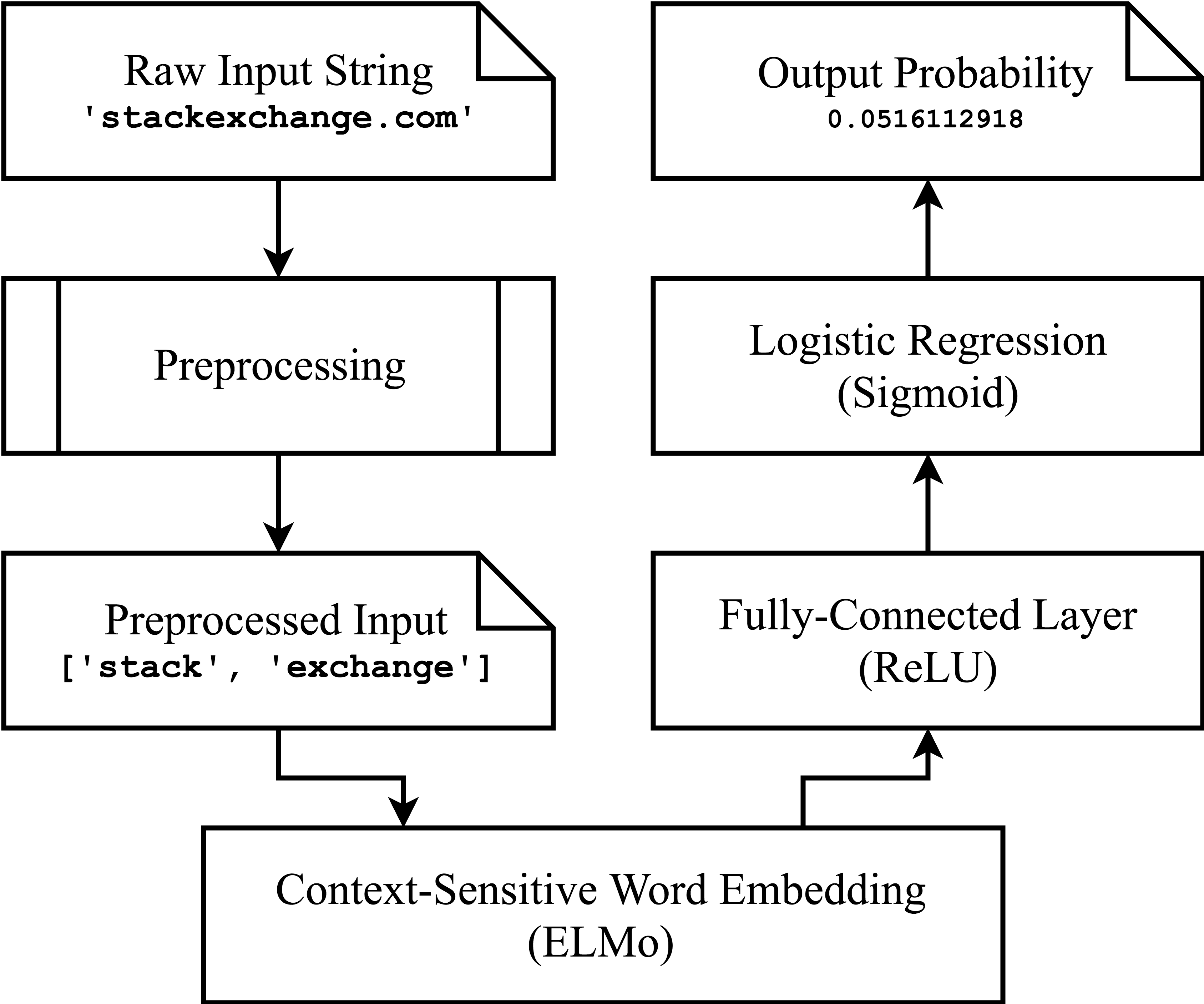}}
\caption{The model architecture.}
\label{architecture}
\end{figure}

The pre-trained ELMo embedding was loaded using TensorFlow Hub. This version of the embedding was trained by Google on the 1 Billion Word Benchmark dataset \cite{chelba2014}. We used the \verb|default| key of the module's output dictionary to obtain a fixed mean-pooling of the contextualized word representations generated by the model \cite{tfhub-elmo}.

The rest of the model comprised a fully-connected layer with 128 rectified linear units (ReLUs) and a logistic regression output layer. We did not incorporate dropout due to there having been no indications of decreased overfitting with it employed. The overall model architecture is shown in Fig.~\ref{architecture}.

\section{Experiments}

\subsection{Evaluation Metrics}

We evaluated the performance of the models with several metrics. The true positive rate (TPR):
\begin{equation}
\text{TPR} = \frac{\sum \text{True Positive}}{\sum \text{Condition Positive}}\text{,}\label{tpr}
\end{equation}
and false positive rate (FPR):
\begin{equation}
\text{FPR} = \frac{\sum \text{False Positive}}{\sum \text{Condition Negative}}\text{,}\label{fpr}
\end{equation}
were evaluated at different classifier discrimination thresholds to plot receiver operating characteristic (ROC) curves. These curves are useful due to their relative insensitivity to class imbalance \cite{fawcett2006}. The area under the ROC curve is known as the AUC, and is equal to the probability that a random positive instance would be ranked higher than a random negative instance by the classifier, where rank ordering is determined by the output probability predicted by the classifier \cite{fawcett2006}. Since inline DGA detection requires low false positive rates to be feasible for real-world deployment, we were particularly interested in the partial AUC, which is the standardized AUC for a given FPR ceiling \cite{mcclish1989}. For our purposes, we set the FPR ceiling at 1\%. With the same considerations, we also looked at the TPR when the FPR is at or below 1\% and 0.1\%.

Finally, we considered precision, defined as
\begin{equation}
\text{Precision} = \frac{\sum \text{True Positive}}{\sum \text{Predicted Condition Positive}}\label{precision},
\end{equation}
and recall, which is synonymous with the TPR as previously defined in \eqref{tpr}. We additionally looked at the harmonic mean of recall and precision, known as the $F_1$ score:
\begin{equation}
F_1 = 2 \cdot \frac{\text{Precision} \cdot \text{Recall}}{\text{Precision} + \text{Recall}}\label{f1}.
\end{equation}
The $F_1$ score is a measure of the relevance of the classifier's labeling, both in terms of the probability that a positively labeled instance is relevant, as well as the probability that a relevant instance is labeled positive.

\subsection{Experimental Design}

The benign and DGA domains used in the experiments were from the following sources:
\begin{itemize}
\item OpenDNS public domain lists\footnote{https://github.com/opendns/public-domain-lists} provided 20,000 instances of benign domains, consisting of a random sample of 10,000 domains as well as the top 10,000 domains with the most DNS queries (in November 2014);
\item Andrey Abakumov's repository\footnote{https://github.com/andrewaeva/DGA} provided the algorithms and wordlists for \verb|matsnu| and \verb|rovnix|;
\item Johannes Bader's repository\footnote{https://github.com/baderj/domain\_generation\_algorithms} provided the algorithm and wordlists for \verb|pizd| and \verb|suppobox|.
\end{itemize}

The technique was evaluated with the following experiments:
\begin{enumerate}
\item models trained individually on single-DGA datasets of varying sizes, to measure the models' ability to learn from limited data,
\item models trained on multiple-DGA datasets of varying sizes, to test the models' capacity to simultaneously learn multiple DGA families.
\end{enumerate}

The primary aim of the first experiment was to assess the binary decision process of the classifiers when trained on small amounts of data. Each DGA was tested separately, and metrics were reported by class.

Considering that a detection system deployed in the real-world would likely have to contend with domains from multiple DGAs, the second experiment was meant to evaluate the model's performance in a multiple-DGA scenario, and discover the degree of performance degradation when multiple DGAs are involved.

In all experiments, the training dataset contained 14,000 benign domains, and the disjoint validation and test datasets each contained 3,000 benign domains. Likewise for DGA domains, each DGA that the model was to be tested against had 1,000 domains included in the test dataset that would never be used in training or validation. Of the DGA domains that would be seen in the training process, we used a 80/20 split for the training and validation datasets, i.e. if there were 120 observations to be used for the training process, 96 were included in the training dataset and the remaining 24 were placed in the validation dataset.

\section{Results}

\begin{table*}[htbp]
\begin{center}
\caption{First experiment: results for models trained on single-DGA datasets}
\label{results-1}
\begin{tabular}{ccccccccccc}
\hline
DGA & Observations & Training & Validation & Precision & Recall & $F_1$ Score & Partial AUC & AUC & $\text{TPR}_{\text{FPR} < 0.01}$ & $\text{TPR}_{\text{FPR} < 0.001}$ \\
\hline
                    & 30        & 24        & 6         & 1.0000    & 0.4440    & 0.6150    & 0.9709    & 0.9986    & 0.9660    & 0.8950    \\
                    & 60        & 48        & 12        & 1.0000    & 0.6870    & 0.8145    & 0.9751    & 0.9988    & 0.9800    & 0.8860    \\
                    & 90        & 72        & 18        & 1.0000    & 0.7860    & 0.8802    & 0.9824    & 0.9993    & 0.9880    & 0.9280    \\
\verb|matsnu|
                    & 120       & 96        & 24        & 1.0000    & 0.9010    & 0.9479    & 0.9893    & 0.9995    & 0.9890    & 0.9570    \\
                    & 240       & 192       & 48        & 0.9978    & 0.9180    & 0.9563    & 0.9876    & 0.9996    & 0.9930    & 0.9180    \\
                    & 480       & 384       & 96        & 0.9989    & 0.9500    & 0.9739    & 0.9889    & 0.9996    & 0.9930    & 0.9660    \\
                    & 960       & 768       & 192       & 0.9979    & 0.9640    & 0.9807    & 0.9932    & 0.9998    & 0.9950    & 0.9700    \\
\hline
                    & 30        & 24        & 6         & 1.0000    & 0.4870    & 0.6550    & 0.9482    & 0.9963    & 0.9380    & 0.8090    \\
                    & 60        & 48        & 12        & 1.0000    & 0.6470    & 0.7857    & 0.9601    & 0.9977    & 0.9560    & 0.8240    \\
                    & 90        & 72        & 18        & 1.0000    & 0.7890    & 0.8877    & 0.9723    & 0.9989    & 0.9740    & 0.9170    \\
\verb|rovnix|
                    & 120       & 96        & 24        & 1.0000    & 0.7610    & 0.8643    & 0.9695    & 0.9987    & 0.9610    & 0.9100    \\
                    & 240       & 192       & 48        & 0.9988    & 0.8520    & 0.9196    & 0.9813    & 0.9994    & 0.9860    & 0.9010    \\
                    & 480       & 384       & 96        & 0.9968    & 0.9310    & 0.9628    & 0.9863    & 0.9995    & 0.9870    & 0.9240    \\
                    & 960       & 768       & 192       & 0.9928    & 0.9670    & 0.9797    & 0.9877    & 0.9997    & 0.9930    & 0.9550    \\
\hline
                    & 30        & 24        & 6         & 0.9960    & 0.2510    & 0.4010    & 0.7843    & 0.9690    & 0.6890    & 0.3470    \\
                    & 60        & 48        & 12        & 0.9957    & 0.4610    & 0.6302    & 0.8361    & 0.9873    & 0.7690    & 0.5560    \\
                    & 90        & 72        & 18        & 0.9984    & 0.6410    & 0.7808    & 0.9198    & 0.9958    & 0.9160    & 0.7490    \\
\verb|pizd|
                    & 120       & 96        & 24        & 0.9948    & 0.7700    & 0.8681    & 0.9102    & 0.9953    & 0.9110    & 0.6180    \\
                    & 240       & 192       & 48        & 0.9977    & 0.8780    & 0.9340    & 0.9734    & 0.9992    & 0.9860    & 0.9110    \\
                    & 480       & 384       & 96        & 0.9900    & 0.9880    & 0.9890    & 0.9900    & 0.9996    & 0.9980    & 0.9380    \\
                    & 960       & 768       & 192       & 0.9841    & 0.9890    & 0.9865    & 0.9586    & 0.9991    & 0.9980    & 0.6970    \\
\hline
                    & 30        & 24        & 6         & 1.0000    & 0.2030    & 0.3375    & 0.7635    & 0.9563    & 0.6190    & 0.3490    \\
                    & 60        & 48        & 12        & 0.9933    & 0.4480    & 0.6175    & 0.8379    & 0.9790    & 0.7690    & 0.3900    \\
                    & 90        & 72        & 18        & 0.9962    & 0.5220    & 0.6850    & 0.8586    & 0.9872    & 0.8330    & 0.5950    \\
\verb|suppobox|
                    & 120       & 96        & 24        & 0.9855    & 0.6810    & 0.8054    & 0.8551    & 0.9870    & 0.8030    & 0.5860    \\
                    & 240       & 192       & 48        & 0.9828    & 0.8560    & 0.9150    & 0.9112    & 0.9975    & 0.9420    & 0.6060    \\
                    & 480       & 384       & 96        & 0.9936    & 0.9340    & 0.9629    & 0.9653    & 0.9990    & 0.9830    & 0.7620    \\
                    & 960       & 768       & 192       & 0.9919    & 0.9770    & 0.9844    & 0.9887    & 0.9998    & 0.9980    & 0.8980    \\
\hline
\end{tabular}
\end{center}
\end{table*}

\begin{table*}[htbp]
\begin{center}
\caption{Second experiment: results for models trained on multiple-DGA datasets}
\label{results-2}
\begin{tabular}{ccccccccccc}
\hline
DGA & Observations & Training & Validation & Precision & Recall & $F_1$ Score & Partial AUC & AUC & $\text{TPR}_{\text{FPR} < 0.01}$ & $\text{TPR}_{\text{FPR} < 0.001}$ \\
\hline
\verb|matsnu|       & 100       & 80        & 20        & 0.9732    & 0.8730    & 0.9204    & 0.8768    & 0.9941    & 0.8900    & 0.5410    \\
\verb|rovnix|       & 100       & 80        & 20        & 0.9739    & 0.8970    & 0.9339    & 0.8992    & 0.9954    & 0.9090    & 0.6250    \\
\verb|pizd|         & 100       & 80        & 20        & 0.9727    & 0.8560    & 0.9106    & 0.8554    & 0.9934    & 0.8770    & 0.4620    \\
\verb|suppobox|     & 100       & 80        & 20        & 0.9708    & 0.7980    & 0.8760    & 0.8303    & 0.9898    & 0.8160    & 0.4130    \\
\hdashline
Micro Average       & 400       & 320       & 80        & 0.9930    & 0.8560    & 0.9194    & 0.8656    & 0.9932    & 0.8733    & 0.5108    \\
\hline
\verb|matsnu|       & 200       & 160       & 40        & 0.9874    & 0.8620    & 0.9204    & 0.9183    & 0.9968    & 0.9340    & 0.6380    \\
\verb|rovnix|       & 200       & 160       & 40        & 0.9871    & 0.8420    & 0.9088    & 0.9122    & 0.9965    & 0.9150    & 0.6540    \\
\verb|pizd|         & 200       & 160       & 40        & 0.9876    & 0.8780    & 0.9296    & 0.9255    & 0.9970    & 0.9530    & 0.6450    \\
\verb|suppobox|     & 200       & 160       & 40        & 0.9874    & 0.8630    & 0.9210    & 0.9123    & 0.9967    & 0.9260    & 0.6010    \\
\hdashline
Micro Average       & 800       & 640       & 160       & 0.9968    & 0.8608    & 0.9238    & 0.9171    & 0.9967    & 0.9313    & 0.6338    \\ 
\hline
\verb|matsnu|       & 400       & 320       & 80        & 0.9361    & 0.9520    & 0.9440    & 0.8859    & 0.9956    & 0.8920    & 0.5920    \\
\verb|rovnix|       & 400       & 320       & 80        & 0.9359    & 0.9490    & 0.9424    & 0.8830    & 0.9956    & 0.8960    & 0.5510    \\
\verb|pizd|         & 400       & 320       & 80        & 0.9375    & 0.9750    & 0.9559    & 0.8558    & 0.9960    & 0.8900    & 0.4040    \\
\verb|suppobox|     & 400       & 320       & 80        & 0.9376    & 0.9770    & 0.9569    & 0.8550    & 0.9959    & 0.8990    & 0.3810    \\
\hdashline
Micro Average       & 1600      & 1280      & 320       & 0.9834    & 0.9633    & 0.9732    & 0.8697    & 0.9958    & 0.8933    & 0.4800    \\
\hline
\end{tabular}
\end{center}
\end{table*}

\subsection{Training \& Evaluation Durations}

A key advantage of our technique is the considerable reduction in training durations. Due to the small number of trainable parameters (comparable with the LSTM model in \cite{woodbridge2016}) and minimal training data required, the models could be trained extremely quickly. On a consumer-grade system with two \textit{NVIDIA GeForce GTX 1080 Ti} GPUs, each training epoch took less than a minute. Typically, fewer than 10 epochs were required for convergence. Inference on the same system took 3 milliseconds per domain.

\begin{figure}[htbp]
\centering{\includegraphics[width=0.95\linewidth]{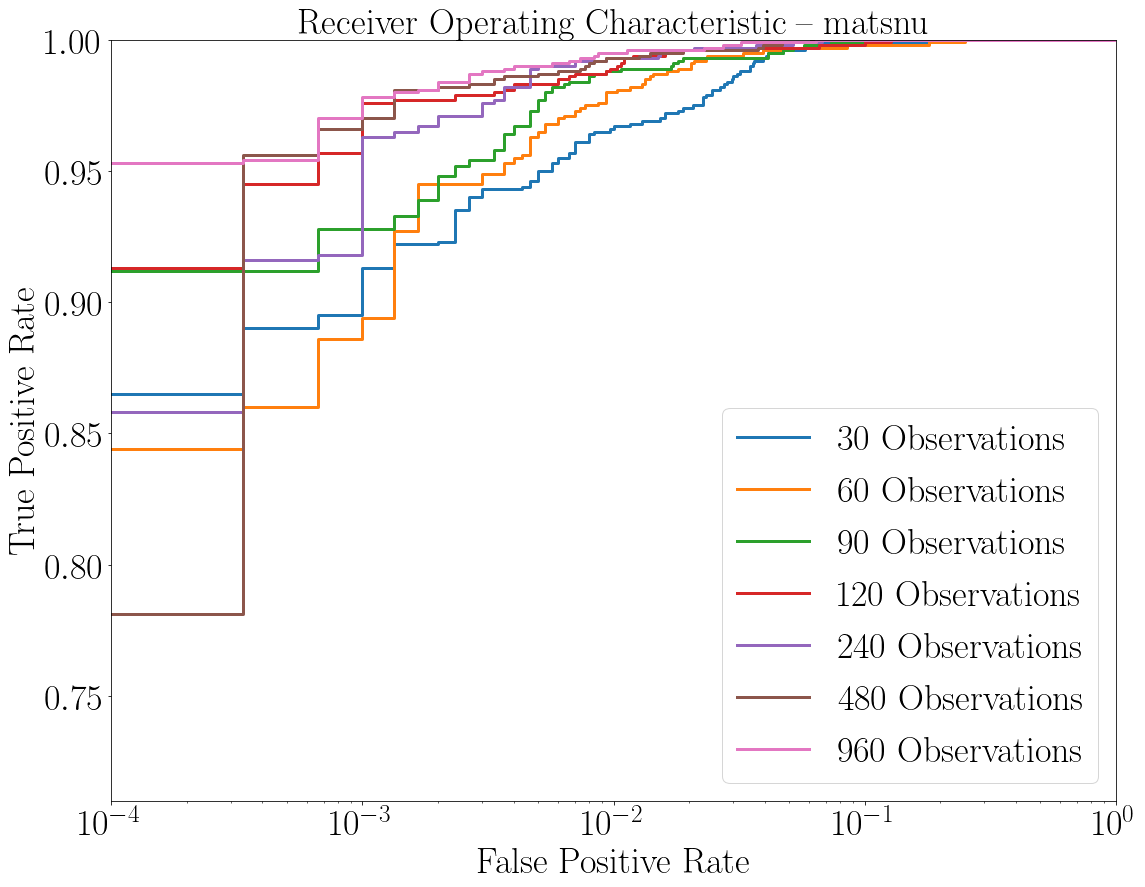}}
\cprotect\caption{ROC curves of the \verb|matsnu| classifier for different observation counts (single-DGA models). Note that the y-axis is on a different scale compared to Fig~\ref{ed-1-roc-pizd}.}
\label{ed-1-roc-matsnu}
\end{figure}

\begin{figure}[htbp]
\centering{\includegraphics[width=0.95\linewidth]{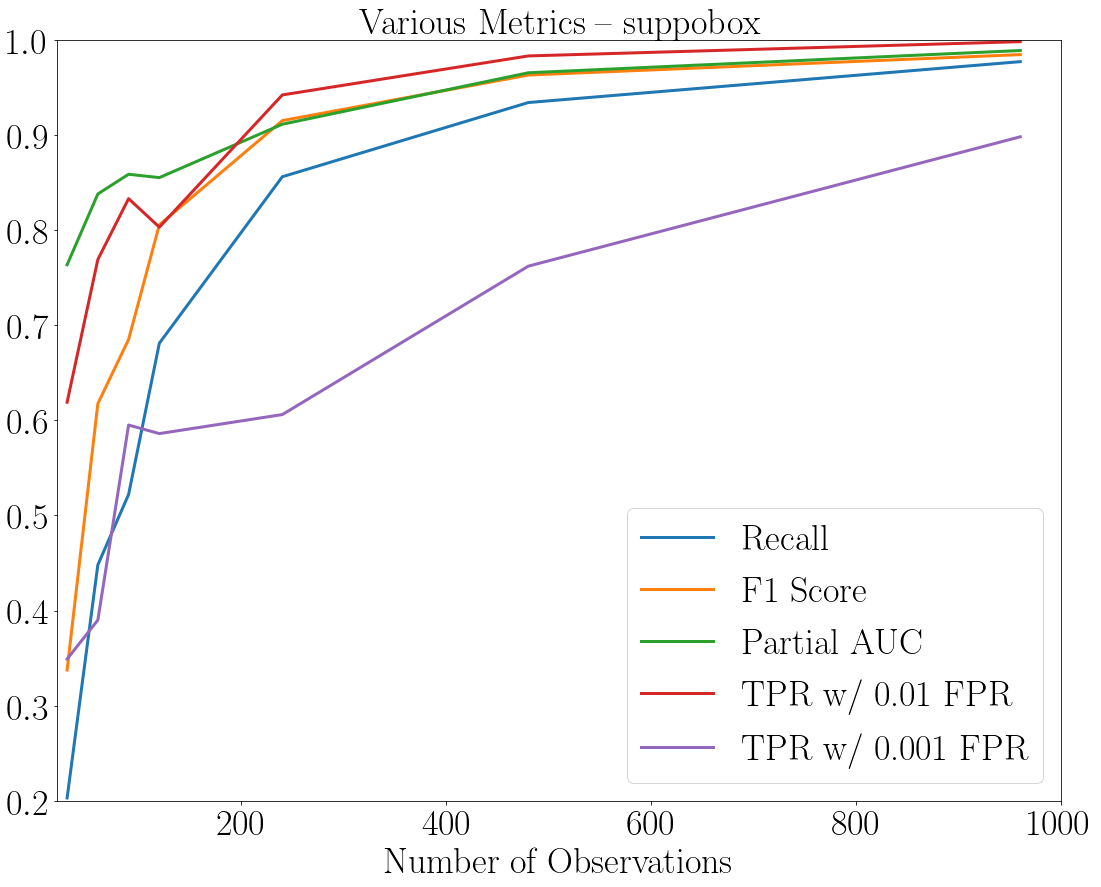}}
\cprotect\caption{Metrics for the \verb|suppobox| classifier with increasing observation count (single-DGA models).}
\label{ed-1-metrics-suppobox}
\end{figure}

\begin{figure}[htbp]
\centering{\includegraphics[width=0.95\linewidth]{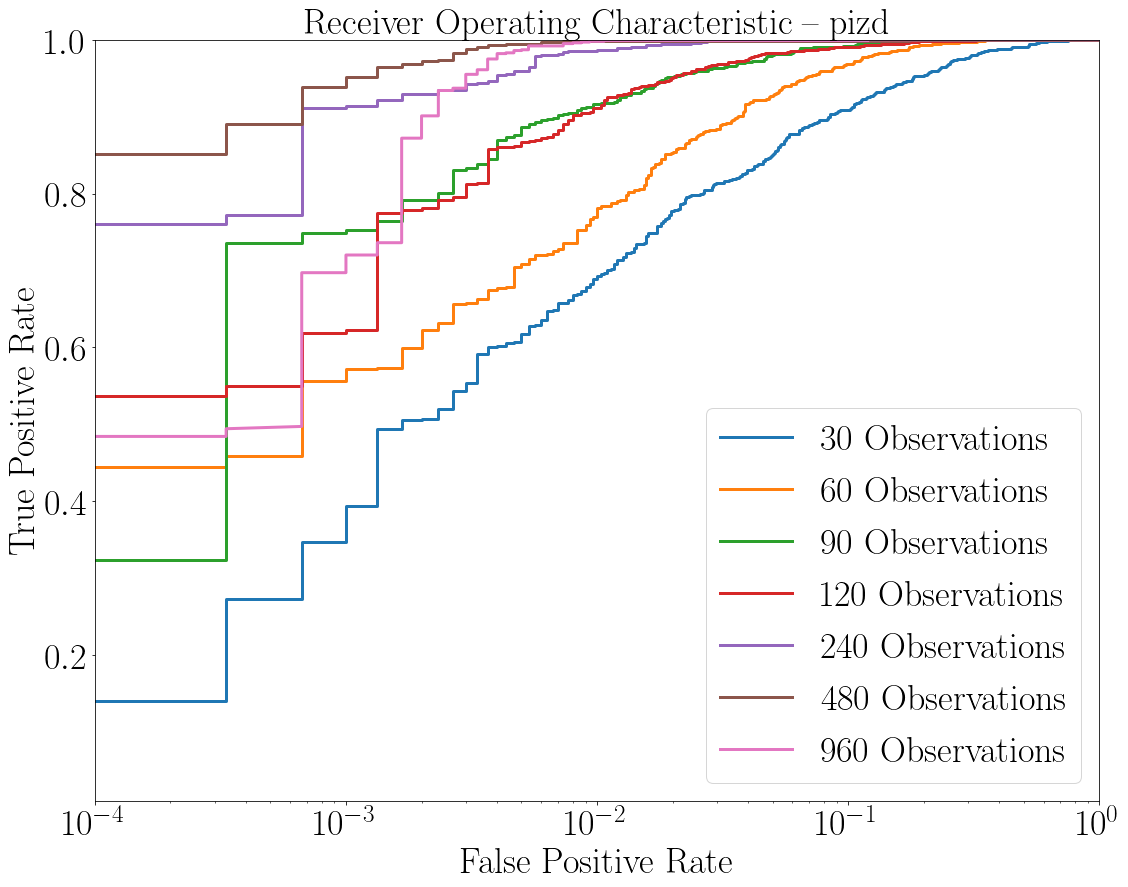}}
\cprotect\caption{ROC curves of the \verb|pizd| classifier for different observation counts (single-DGA models). Note that the y-axis is on a different scale compared to Fig~\ref{ed-1-roc-matsnu}.}
\label{ed-1-roc-pizd}
\end{figure}

\begin{figure}[htbp]
\centering{\includegraphics[width=0.95\linewidth]{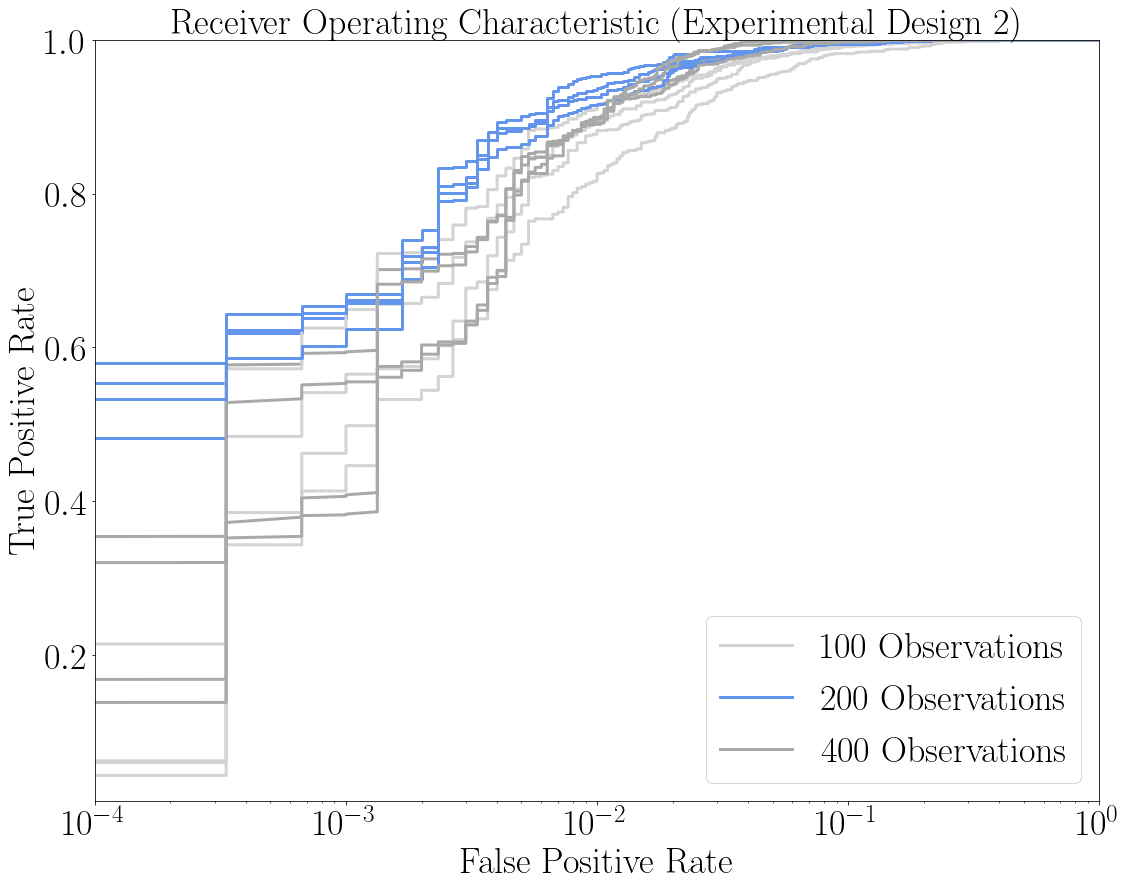}}
\caption{ROC curves for the second experiment (multiple-DGA models).}
\label{ed-2-all}
\end{figure}

\subsection{Single-DGA Experiment}

The results for the first experiment are given in Table~\ref{results-1}. We see that the models performed exceedingly well even with just 30 observations available to the training process, especially for the \verb|matsnu| and \verb|rovnix| classifiers. The ROC curve for the \verb|matsnu| classifier is given in Fig.~\ref{ed-1-roc-matsnu}. Though it is not reflected in Table~\ref{results-1}, it is of note that the \verb|matsnu| classifier was capable of obtaining a 0.9120 TPR at an FPR of 0.0001 with only 90 observations used in training.

The classifiers for the other two DGAs required more training observations to attain peak performance, though we began to encounter diminishing returns past 480 observations (see Fig.~\ref{ed-1-metrics-suppobox} for the impact of increasing the number of observations on the various metrics for the \verb|suppobox| classifier). Interestingly, the \verb|pizd| classifier actually performed worse when we increased the number of observations from 480 to 960. This can be seen in the ROC curve (Fig.~\ref{ed-1-roc-pizd}); if we compare the brown (480 observations) and pink lines (960 observations), we see that the TPR for the lower range of FPR suffered tremendously with the increase in observations.

\subsection{Multiple-DGA Experiment}

The results for the second experiment are given in Table~\ref{results-2}. We see that the performance, while inferior to the single-DGA setup, was still excellent and comparable to other techniques in the literature. There also seems to be a performance sweet spot at 200 observations per DGA. This is also visible in the ROC plot (Fig.~\ref{ed-2-all}), where we see that the curves for all four DGAs were lower for 100 and 400 observations.

\section{Conclusion}

We presented a technique for performing domain classification based on word-level information to detect wordlist-based DGA domains. This was done by using a pre-trained context-sensitive word embedding (specifically ELMo \cite{peters2018}) with a simple classification network. As supported by our results, we were able to take advantage of the language semantics knowledge stored by ELMo to achieve state-of-the-art wordlist-based DGA domain detection with minimal training data. The small amount of training data required for the technique is especially noteworthy for defending against malware families that generate small numbers of domains per day (e.g. \verb|matsnu| generates only 10 domains daily \cite{skuratovich2015}).

Future work could focus on further developing the embedding process due to its outsized contribution to the performance of the architecture. Though the unmodified pre-trained ELMo representations performed extremely well in our experiments, there might be room for improvement. Being generated from a biLM, the embedding could be amenable to fine-tuning via techniques like Universal Language Model Fine-tuning for Text Classification (ULMFiT) \cite{howard2018} to enhance its effectiveness as a component in this DGA detection technique. We also intend to investigate the architecture's performance on DGAs that have not been previously encountered.

\section*{Acknowledgment}

The authors would like to thank Tim Strawbridge and the anonymous reviewers for their invaluable comments and suggestions.

\printbibliography

\end{document}